\documentclass[aps,pra,reprint,amsmath,amssymb,superscriptaddress,nobibnotes, longbibliography]{revtex4-1}
\usepackage{graphicx,SIunits}
\usepackage{bm}     
\usepackage{hyperref} 
\usepackage{dsfont}    
\usepackage{color}

\usepackage{amsmath}
\usepackage{amsthm}
\usepackage{amssymb}
\usepackage{tikz}

\usetikzlibrary{decorations.pathreplacing}
\usepackage{braket}
\usepackage{dsfont}

\usepackage{color}
\usepackage{lineno}
\usepackage{etoolbox}

\begin{document}

\preprint{APS/123-QED}

\title{Contextuality-enhanced quantum state discrimination under fixed failure probability}

  \author{Min Namkung}%
      \affiliation{Center for Quantum Technology, Korea Institute of Science and Technology (KIST), Seoul 02792, Korea}
      
  \author{Hyang-Tag Lim}
      \email{hyangtag.lim@kist.re.kr}
      \affiliation{Center for Quantum Technology, Korea Institute of Science and Technology (KIST), Seoul 02792, Korea}
      \affiliation{Division of Quantum Information, KIST School, Korea University of Science and Technology, Seoul 02792, Korea}

\date{\today}


\begin{abstract}
Quantum state discrimination enables the accurate identification of quantum states, which are generally nonorthogonal. Among various strategies, minimum-error discrimination and unambiguous state discrimination exhibit contextuality-enhanced success probabilities that surpass classical bounds, offering significant advantages for quantum sensing and communication. However, in practice, both error and failure outcomes can occur, suggesting the need for a unified strategy that incorporates both aspects while exploring the potential for contextuality enhancement. In this work, we theoretically demonstrate contextuality enhancement in quantum state discrimination under a fixed failure probability. We show that this enhancement disappears within a certain intermediate range of failure probabilities--a phenomenon absent in conventional strategies, where both minimum-error and unambiguous discrimination consistently outperform the noncontextual bound for equal priors.
{Moreover}, we analyze how the existence of this non-enhancement region depends on the confusability of the quantum states, which corresponds to their fidelity in a quantum model. {We further extend the discussion to the noisy state discrimination, which even encompasses the maximal-confidence discrimination. In this extended discussion, we observe that the non-enhancement region tends to disappear with increasing noise strength.}
\end{abstract}


\maketitle

\section{Introduction}
A quantum system is generally described by nonorthogonal quantum states~\cite{c.w.helstrom,a.s.holevo,j.a.bergou,s.m.barnett,j.bae}, making it crucial to develop measurement strategies that identify such states with the highest possible success probability. This necessity highlights the significance of quantum state discrimination for understanding quantum systems, ultimately contributing to the advancement of quantum information science. Beyond foundational studies, quantum state discrimination plays a central role in various prepare-and-measure tasks, including quantum sensing~\cite{h.yuan,m.hillery,n.ali}, quantum communication~\cite{i.a.burenkov,g.cariolaro}, and quantum computation~\cite{x.-f.yin,j.tilly}. In principle, quantum states can be discriminated without failure using minimum-error discrimination~\cite{y.c.eldar,j.bae2,d.ha,m.namkung}, or without error using unambiguous state discrimination~\cite{i.d.ivanovic,d.dieks,a.peres,g.jaeger,j.a.bergou2,d.ha2}. However, in realistic settings, both error and failure outcomes inevitably occur with non-zero probabilities. This motivates a unified approach that bridges these two strategies of quantum state discrimination with fixed failure probability~\cite{a.chefles,c.w.zhang,j.fiurasek,u.herzog,d.fields}. An advantage of this generalized strategy is that it can be experimentally tuned by adjusting the allowable failure probability~\cite{s.gomez}, and thus encompassing the broadest class of measurement configurations.

A particularly intriguing aspect of quantum state discrimination is its connection to quantum contextuality~\cite{d.schmid,j.shin,s.mukerjee,j.shin2,k.flatt,k.flatt2,c.r.i.carceller_}. In the noncontextual models proposed by Kochen, Specker, and Spekkens~\cite{c.budroni,s.kochen,r.w.spekkens}, measurement outcomes are determined by hidden variables, independent of quantum postulates, and applicable to arbitrary quantum states. Consequently, violations of such models expand the notion of nonclassicality, complementing quantum nonlocality defined in multipartite systems~\cite{j.s.bell,j.f.clauser}. It has recently been shown that noncontextual models fail to reproduce the success probabilities achieved by both minimum-error~\cite{d.schmid,s.mukerjee,j.shin} and optimal unambiguous~\cite{j.shin2,k.flatt} discrimination strategies. Given that these strategies are foundational for numerous quantum information protocols~\cite{i.a.burenkov,g.cariolaro,l.-m.duan,m.lostaglio,j.b.brask,c.r.i.carceller,v.buzek}, their contextuality enhancement underscores a vital advantage for emerging quantum technologies. Specifically, when two pure states are prepared with equal prior probabilities, both strategies consistently outperform the noncontextual bound~\cite{d.schmid,k.flatt}. Given the practical relevance of the unified quantum state discrimination strategy, it is crucial to address the following question: Can contextuality-enhanced success probability be achieved for all values of the fixed failure probability?

In this work, we investigate contextuality enhancement in this generalized quantum state discrimination framework. We first analytically derive the noncontextual upper bound of the success probability, applicable to arbitrary values of failure probability, prior probabilities, and pure state configurations. Our analysis reveals the existence of a specific range of failure probabilities where contextuality enhancement is not achievable--even when the prior probabilities are equal. This contrasts with both the minimum-error and unambiguous strategies, which always violate the noncontextual bound under equal priors. Moreover, we show that this non-enhancement region depends on the confusability between the two quantum states--i.e., how indistinguishable they are--which is mathematically equivalent to the quantum fidelity~\cite{r.josza,m.a.nielson}. As confusability increases, contextuality enhancement becomes more likely at higher failure probabilities, and vice versa. {Besides, as a quantum system is noisy in general, we further consider discrimination between mixed states, which even covers the maximal-confidence discrimination~\cite{s.croke,k.flatt,k.flatt2}. We observe that the contextuality enhancement tends to be pronounced with increasing noise strength.}

\section{Background}

\subsection{Scenario}
\begin{figure}[t]
\centerline{\includegraphics[width=8.2cm]{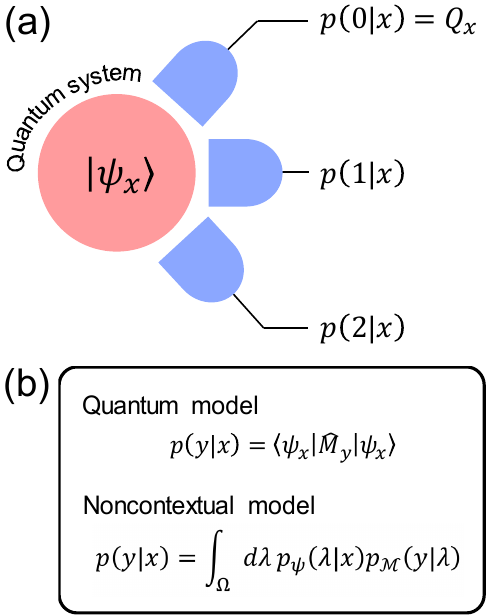}}
\caption{(a) Strategy for discriminating one of two quantum states $|\psi_1\rangle$ and $|\psi_2\rangle$, prepared with prior probabilities $q_1$ and $q_2$, respectively. Here, a measurement for discriminating these quantum states is established to output one of three outcomes $y=0,1,2$. If $y=1,2$, then one can identify the prepared state as $|\psi_y\rangle$, although incorrect sometimes. Otherwise, i.e., $y=0$, it means that the identification is failed. In the strategy, the probability $p(0|x)$ is fixed to a constant $Q_x$, leading to the fixed failure probability $Q=q_1Q_1+q_2Q_2$. (b) Representation of measurement probability. In a quantum model, conditional probability to obtain $y$ for given $|\psi_x\rangle$ is represented by $p(y|x)=\langle\psi_x|\hat{M}_y|\psi_x\rangle$. Meanwhile, in a noncontextual model, this conditional probability represented in terms of a hidden variable $\lambda$, based on the Bayesian theorem.}
\centering
\label{fig:1}
\end{figure}

We begin by introducing the scenario of quantum state discrimination, as illustrated in in Fig.~\ref{fig:1}. A quantum system is described by two nonorthogonal quantum states, $|\psi_1\rangle$ and $|\psi_2\rangle$, prepared with prior probabilities $q_1$ and $q_2$, respectively. To identify the state of the quantum system, a measurement represented by a positive operator-valued measure (POVM) $\mathcal{M}=\{\hat{M}_0,\hat{M}_1,\hat{M}_2\}$, where $\hat{M}_y$ is the POVM element corresponding to an outcome $y$. If $y\not=0$ is obtained with the probability $p(y|x)=\langle\psi_x|\hat{M}_y|\psi_x\rangle$, then the measurement directs the quantum state to $|\psi_y\rangle$ with nonzero correct probability, and the state discrimination is failed otherwise. In the quantum model, the success probability of discriminating the quantum states is formulated according to the Born's rule as
\begin{eqnarray}
    P_{\rm succ.}^{\rm (Q)}=q_1\langle\psi_1|\hat{M}_1|\psi_1\rangle+q_2\langle\psi_2|\hat{M}_2|\psi_2\rangle.
\end{eqnarray}
In the state discrimination of this work, it is assumed that the failure probability is fixed to a constant, denoted by
\begin{eqnarray}
    Q=q_1\langle\psi_1|\hat{M}_0|\psi_1\rangle+q_2\langle\psi_2|\hat{M}_0|\psi_2\rangle.
\end{eqnarray}
In other words, the optimal strategy for maximizing the success probability is described by the solution of the following optimization problem:
\begin{eqnarray}\label{opt_prob}
    \mathrm{maximize} & \  & P_{\rm succ.}^{\rm (Q)} \nonumber\\
    \mathrm{subject \ to} & \ & \hat{M}_y\ge0, \ \hat{M}_0+\hat{M}_1+\hat{M}_2=\mathbb{I}, \nonumber\\
    && Q=\mathrm{constant},
\end{eqnarray}
with the identity operator $\mathbb{I}$. When $Q$ is equal to zero, the above optimization leads to the minimum-error discrimination, and it also covers the optimal unambiguous state discrimination for large $Q$. It was shown that the maximum success probability is analytically given by
\begin{eqnarray}\label{p_q}
    \max_{\mathcal{M}}P_{\rm succ.}^{\rm (Q)}=\frac{\bar{Q}{+}\left\{\bar{Q}^2-(2\sqrt{q_1q_2c_{\psi_1,\psi_2}}-Q)^2\right\}^{\frac{1}{2}}}{2},
\end{eqnarray}
with the confusability $c_{\psi_1,\psi_2}=|\langle\psi_1|\psi_2\rangle|^2$ and $\bar{Q}=1-Q$, if $Q\le 2\sqrt{q_1q_2c_{\psi_1,\psi_2}}$, and 
\begin{eqnarray}\label{p_q2}
    \max_{\mathcal{M}}P_{\rm succ.}^{\rm (Q)}=1-Q,
\end{eqnarray}
otherwise~{\cite{u.herzog,d.fields,s.gomez}}. Note that the critical value is equal to the minimum failure probability of the unambiguous state discrimination~\cite{i.d.ivanovic,d.dieks,a.peres,g.jaeger}. It means that the optimal strategy is to perform the optimal unambiguous state discrimination when $Q\ge2\sqrt{q_1q_2c_{\psi_1,\psi_2}}$, leading to $\max_{\mathcal{M}}P_{\rm succ.}^{\rm (Q)}=1-2\sqrt{q_1q_2c_{\psi_1,\psi_2}}$ for large $Q$.

{We emphasize that the criterion of the theorem is particularly advantageous for practical quantum task, in which a measurement may have imperfection. For example, let us consider discriminating two polarization states of a single photon. It is rational in the practical manner to consider an inefficient photodetector~\cite{g.cariolaro}, resulting in photon loss probability $Q$. In this case, adjusting the measurement configuration is exactly same as what is exactly described in Eq.~(\ref{opt_prob}).}

\subsection{Noncontextual model}
{In the noncontextual model, a measurement probability is represented in terms of a hidden variable, without employing postulates of quantum measurement theory. Specifically, the probability to obtain an outcome $y$ for given label $x$ of the prepared system is formulated as~\cite{s.kochen,r.w.spekkens,d.schmid,s.mukerjee,j.shin,j.shin2,k.flatt}
\begin{eqnarray}
    p(y|x)=\int_{\Omega}d\lambda p_{\psi}(\lambda|x)p_{\mathcal{M}}(y|\lambda),
\end{eqnarray}
based on the Bayesian theorem. Here, $\Omega$ denotes a set of the hidden variables, $p_{\psi}(\lambda|x)$ represents the prepared system, and $p_{\mathcal{M}}(y|\lambda)$ are response functions constituting a measurement. Both two functions are non-negative, satisfying 
\begin{eqnarray}
    \int_{\Omega}d\lambda p_{\psi}(\lambda|x)=1, & & \mathrm{for \ all \ } x=1,2, \nonumber\\
        \sum_{y=0}^{2}p_{\mathcal{M}}(y|\lambda)=1, & & \mathrm{for \ all \ } \lambda\in\Omega.
\end{eqnarray}
Based on this formulation, the success probability of the discrimination is described by
\begin{eqnarray}\label{p_nc_back}
    P_{\rm succ.}^{\rm (NC)}&=&q_1\int_{\Omega}d\lambda p_{\psi}(\lambda|1)p_{\mathcal{M}}(1|\lambda)\nonumber\\
    &+&q_2\int_{\Omega}d\lambda p_{\psi}(\lambda|2)p_{\mathcal{M}}(2|\lambda),
\end{eqnarray}
and the fixed failure probability is 
\begin{eqnarray}\label{q}
    Q&=&q_1\int_{\Omega}d\lambda p_{\psi}(\lambda|1)p_{\mathcal{M}}(0|\lambda)\nonumber\\
    &+&q_2\int_{\Omega}d\lambda p_{\psi}(\lambda|2)p_{\mathcal{M}}(0|\lambda),
\end{eqnarray}
accordingly. The success probability in Eq.~(\ref{p_nc_back}) is maximized with a certain value of $Q$ in Eq.~(\ref{q}), eventually upper-bounded by a function of $Q$, and thus written as $\bar{P}_{\rm succ.}^{\rm (NC)}(Q)$. If $P_{\rm succ.}^{\rm (Q)}$ can surpass this upper bound, i.e., $P_{\rm succ.}^{\rm (Q)}\ge\bar{P}_{\rm succ.}^{\rm (NC)}$, then it suggests that the quantum state discrimination is contextuality-enhanced, and thus beyond the classical logic.}

\section{Results}

\begin{figure*}[t]
\centerline{\includegraphics[width=18cm]{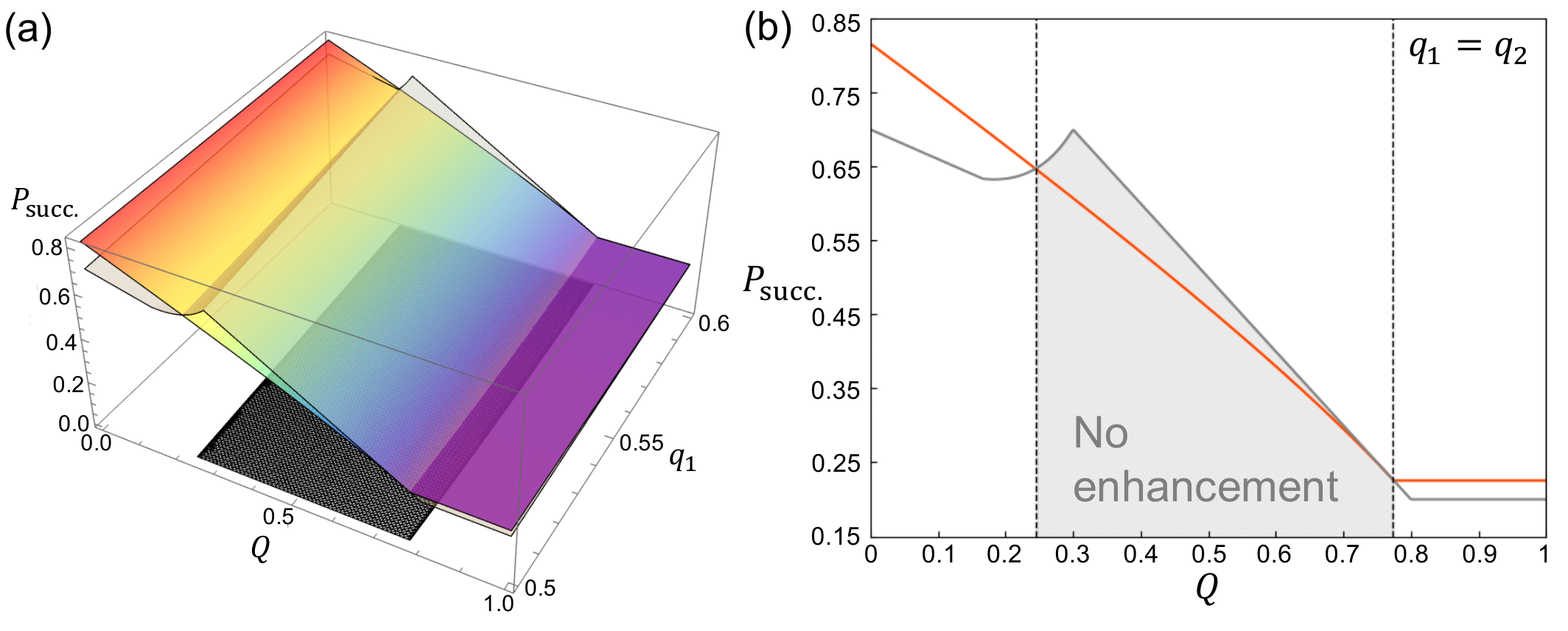}}
\caption{(a) Contextuality-enhanced success probability investigated across $0.5\le q_1\le 0.6$ for the better visibility, given confusability 0.6. Here, the surface colored in rainbow is the maximum success probability under the quantum model, and the gray surface is the noncontextual bound. The shaded area on the 2d plot consisting of $Q$ and $q_1$ axes shows the region where there is {non-enhancement}. (b) The specified {contextuality enhancement} in the case of equal priors. Here, the red (gray) lines show the maximum success probability of the quantum model (the noncontextual bound). The region with {non-enhancement} is observed to $0.245\le Q\le0.7736$.}
\centering
\label{fig:2}
\end{figure*}

\subsection{Upper bound of success probability}
Now, let us theoretically evaluate the noncontextual bound, which is used for demonstrating the {contextuality enhancement}. We first define a function $\Pi(\lambda)=p_{\mathcal{M}}(1|\lambda)+p_{\mathcal{M}}(2|\lambda)$ for simplifying the derivation. Using this function, we rewrite the success probability of Eq.~(\ref{p_nc_back}) as (Detailed information is provided in Appendix A)
\begin{eqnarray}\label{p_nc_rewri}
    P_{\rm succ.}^{\rm (NC)}&=&q_1(1-Q_1)\int_{\Omega}d\lambda p_{\widetilde{\psi}}(\lambda|1)p_{\widetilde{\mathcal{M}}}(1|\lambda)\nonumber\\
    &+&q_2(1-Q_2)\int_{\Omega}d\lambda p_{\widetilde{\psi}}(\lambda|2)p_{\widetilde{\mathcal{M}}}(2|\lambda).
\end{eqnarray}
Here, $Q_x$ are constants denoted by
\begin{eqnarray}
    Q_x=\int_{\Omega}d\lambda p_{\psi}(x|\lambda)p_{\mathcal{M}}(0|\lambda),
\end{eqnarray}
and $p_{\widetilde{\psi}}(\lambda|x)$ and $p_{\widetilde{\mathcal{M}}}(0|\lambda)$ are defined as
\begin{eqnarray}
    p_{\widetilde{\psi}}(\lambda|x)&=&\frac{p_{\psi}(\lambda|x)\Pi(\lambda)}{1-Q_x},\nonumber\\
p_{\widetilde{\mathcal{M}}}(y|\lambda)&=&\frac{1}{\Pi(\lambda)}p_{\mathcal{M}}(y|\lambda),
\end{eqnarray}
respectively. Notably, both two response functions $p_{\widetilde{\mathcal{M}}}(1|\lambda)$ and $p_{\widetilde{\mathcal{M}}}(2|\lambda)$ satisfy $p_{\widetilde{\mathcal{M}}}(1|\lambda)+p_{\widetilde{\mathcal{M}}}(2|\lambda)=1$, and thus the maximization of the success probability in Eq.~(\ref{p_nc_rewri}) is the noncontextual version of the minimum-error discrimination. Therefore, it is possible to use the inequality proven by Refs.~\cite{s.mukerjee,j.shin}, leading to
\begin{eqnarray}\label{p_s_ubb}
    P_{\rm succ.}^{\rm (NC)}\le \bar{Q}-\min\left\{q_1\bar{Q}_1,q_2\bar{Q}_2\right\}c_{\widetilde{\psi}_1,\widetilde{\psi}_2},
\end{eqnarray}
where $\bar{Q}=1-Q$, $\bar{Q}_x=1-Q_x$, and $c_{\widetilde{\psi}_1,\widetilde{\psi}_2}$ is the confusability formulated as
\begin{eqnarray}\label{c}
    c_{\widetilde{\psi}_1,\widetilde{\psi}_2}&=&\int_{\mathrm{supp}[p_{\widetilde{\psi}}(\lambda|2)]}d\lambda p_{\widetilde{\psi}}(\lambda|1)\nonumber\\
    &=&\int_{\mathrm{supp}[p_{\widetilde{\psi}}(\lambda|1)]}d\lambda p_{\widetilde{\psi}}(\lambda|2){,}
\end{eqnarray}
{with the confusability $c_{\psi_1,\psi_2}$ between two prepared states, equals to $|\langle\psi_1|\psi_2\rangle|^2$ of the quantum model~\cite{d.schmid}.} Here, $\mathrm{supp}[p_{\widetilde{\psi}}(\lambda|x)]$ denotes the set of the hidden variables such that $p_{\widetilde{\psi}}(\lambda|x)$ is larger than zero. We can prove that $c_{\widetilde{\psi}_1,\widetilde{\psi}_2}$ satisfies (Detailed derivation is provided in Appendix A)
\begin{eqnarray}\label{c_bound}
    c_{\widetilde{\psi}_1,\widetilde{\psi}_2}\ge\max\left\{\frac{c_{\psi_1,\psi_2}-Q_1}{1-Q_1},\frac{c_{\psi_1,\psi_2}-Q_2}{1-Q_2},0\right\}{.}
\end{eqnarray}
If $c_{\psi_1,\psi_2}$ is less than $Q_x$ for any $x$, then Eq.~(\ref{c_bound}) is simplified to $c_{\widetilde{\psi}_1,\widetilde{\psi}_2}\ge0$. In this case, the upper bound in Eq.~(\ref{p_s_ubb}) becomes $\bar{P}_{\rm succ.}^{\rm (NC)}=1-Q$, which is equal to the maximum success probability of Eq.~(\ref{p_q2}), allowed by the quantum model. This implies that, as $Q$ is rewritten as $Q=q_1Q_1+q_2Q_2$, there is no {contextuality enhancement} in the success probability if $Q$ is selected to be large. Otherwise, i.e., $c_{\psi_1,\psi_2}$ larger than $Q_1$ or $Q_2$, Eq.~(\ref{p_s_ubb}) is rewritten as
\begin{eqnarray}
    P_{\rm succ.}^{\rm (NC)}&\le& f(Q_1,Q_2)\nonumber\\
    &:=&\bar{Q}-\min_x\left\{q_x\bar{Q}_x\right\}\max_y\left\{\frac{c_{\psi_1,\psi_2}-Q_y}{\bar{Q}_y}\right\}.
\end{eqnarray}
The above inequality is satisfied for all possible $Q_1$ and $Q_2$, leading to that the the success probability admits 
\begin{eqnarray}
    P_{\rm succ.}^{\rm (NC)}\le \max_{\substack{0\le Q_1,Q_2\le1 \\ Q=q_1Q_1+q_2Q_2}}f(Q_1,Q_2).
\end{eqnarray}
The maximum value of $f$ can be analytically evaluated as detailed in Appendix B. Through this evaluation, we finally obtain {the following theorem regarding the upper bound of the success probability.}

{\textit{Theorem 1.} For a given fixed failure probability $Q$, the success probability of the noncontextual model in Eq.~(\ref{p_nc_rewri}) is upper-bounded by}
\begin{eqnarray}\label{p_nc_bf}
    \bar{P}_{\rm succ.}^{\rm (NC)}(Q)=\begin{cases}
        \max_{\substack{x=1,2 \\ z\in\left\{0,\frac{Q}{q_x}\right\}}}f_x(z) \ \mathrm{if} \ Q\le q_{\rm min}c_{\psi_1,\psi_2}, \\ 1-Q \ \ \mathrm{otherwise},
    \end{cases}
\end{eqnarray}
where $q_{\rm min}=\min\{q_1,q_2\}$ and $f_x(z)$ is a real function defined as
\begin{eqnarray}
    f_x(z):=\bar{Q}-q_x(1-z)\frac{q_{x\oplus1}c_{\psi_1,\psi_2}-Q+q_xz}{q_{x\oplus1}-Q+q_xz},
\end{eqnarray}
with $x\oplus1=(x\mathrm{ \ mod \ }2)+1$.

{In other words, once an optimal measurement is constructed to exceed the noncontextual bound of Eq.~(\ref{p_nc_bf}), it is useful for revealing the nonclassical aspect of a single system. As this involves the failure probability $Q$, which is mainly associated with a measurement error, it is of critical importance to verify the region of $Q$ that allows the imperfect measurement to still useful for unveiling the nonclassicality.}

\subsection{Verifying contextuality enhancement}

We investigate the behavior analysis of the {contextuality enhancement} as illustrated in Fig.~\ref{fig:2}. In Fig.~\ref{fig:2}(a), the maximum success probability under the quantum model is depicted with varying $Q$ and $q_1$, compared with the noncontextual bound. We note that the case of $Q=0$ corresponds to the minimum-error discrimination, and that of large $Q$ corresponds to the unambiguous state discrimination. Specifically, when $Q$ is larger than a bound $Q_{\rm max}$, it is obvious that the maximum success probability is analytically derived as $1-Q_{\rm max}$, allowed by the optimal unambiguous state discrimination. For example, the maximum success probability is fixed to~\cite{i.d.ivanovic,d.dieks,a.peres,g.jaeger}
\begin{eqnarray}\label{udq}
    P_{\rm succ.}^{\rm (Q)}=1-Q_{\rm max}^{\rm (Q)}=2\sqrt{q_1q_2c_{\psi_1,\psi_2}},
\end{eqnarray}
under the quantum model, and~\cite{j.shin,j.shin2,k.flatt}
\begin{eqnarray}\label{udnc}
    P_{\rm succ.}^{\rm (NC)}=1-Q_{\rm max}^{\rm (NC)}=\max\{q_1,q_2\}c_{\psi_1,\psi_2},
\end{eqnarray}
under the noncontextual model. We observe from Fig.~\ref{fig:2}(a) that the success probability of the quantum model does not surpass the noncontextual bound when $Q$ is within an intermediate range. This suggests that, when interpolating the minimum-error discrimination and the unambiguous state discrimination to cover the practical situation, it is crucial to avoid choosing a failure probability in this intermediate region. 

\begin{figure*}[t]
    \centerline{\includegraphics[width=18cm]{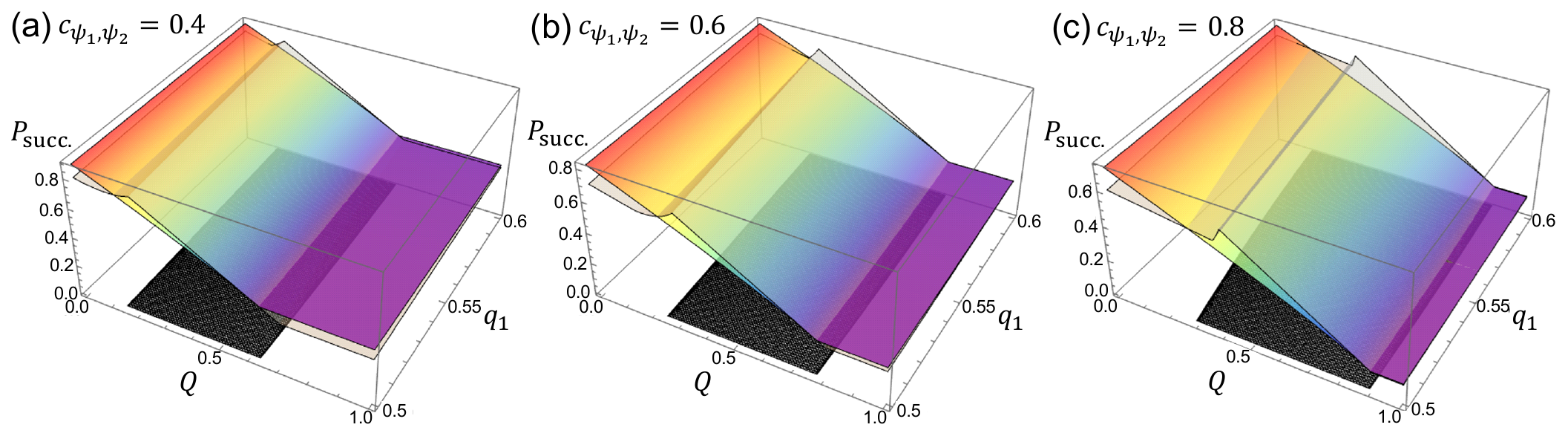}}
    \caption{Contextuality-enhanced success probability investigated across $0.5\le q_1\le 0.6$ for the better visibility. Here, the surface colored in rainbow is the maximum success probability under the quantum model, and the gray surface is the noncontextual bound. The shaded area on the 2d plot consisting of $Q$ and $q_1$ axes shows the region where there is {non-enhancement}. Here, the confusabilities are assumed to be $0.4$, $0.6$, and $0.8$ in (a), (b), and (c), respectively. }
    \centering
    \label{fig:3}
    \end{figure*}

We focus on the case of equal prior probabilities as shown in Fig.~\ref{fig:2}(b). In this case, there is no {contextuality enhancement} in the region of $0.245\le Q\le0.7736$. {Notably, this {non-enhancement} region depends on the prior probabilities. For example, two prior probabilities 0.8 and 0.2 yields the {non-enhancement} region $0.114\le Q\le 0.62$.} We divide the entire region into four parts to display the explicit form the maximum success probability as follows. In the region of $0\le Q\le0.166$, the bound of the noncontextual model is evaluated as
\begin{eqnarray}\label{reg1}
    \bar{P}_{\rm succ.}^{\rm (NC)}=1-\frac{1}{2}c_{\psi_1,\psi_2}-(1-c_{\psi_1,\psi_2})Q,
\end{eqnarray}
becoming equal to the minimum-error discrimination when $Q=0$. In this region, the success probability under the quantum model always outperform the noncontextual bound of Eq.~(\ref{reg1}). In the region of $0.166\le Q\le 0.3$, the noncontextual bound is calculated as
\begin{eqnarray}
    \bar{P}_{\rm succ.}^{\rm (NC)}=1-Q-\frac{1}{2}\frac{c_{\psi_1,\psi_2}-2Q}{1-2Q}.
\end{eqnarray}
Unlike the bound of Eq.~(\ref{reg1}), the above noncontextual bound increases along with $Q$. Due to this behavior, a critical point appears at $Q=0.245$, around which the enhancement is exhibited for $0.166\le Q\le0.245$, while it disappears for $0.245\le Q\le0.3$. In the region of $0.3\le Q\le 0.8$, the noncontextual bound is 
\begin{eqnarray}
    \bar{P}_{\rm succ.}^{\rm (NC)}=1-Q.
\end{eqnarray}
In this region, we observe that the maximal success probability allowed by the quantum theory does not surpass this noncontextual bound if $Q$ is less than or equal to 0.7736. In the region of $Q\ge 0.8$, the maximum success probability under the quantum model is equal to Eq.~(\ref{udq}) if $Q\ge1-\sqrt{c_{\psi_1,\psi_2}}$, and the noncontextual bound is Eq.~(\ref{udnc}) if $Q\ge 1-\frac{1}{2}c_{\psi_1,\psi_2}$. In this region, there exists the {contextuality enhancement} as what is observed from the optimal unambiguous state discrimination.

We further investigate behavior of the region without the {contextuality enhancement} across the confusability, illustrated as Fig.~\ref{fig:3}. In Fig.~\ref{fig:3}, the confusabilities are considered to be 0.4, 0.6, and 0.8, respectively. We observe that, when the confusability is small, the region without the enhancement tends to be placed where the failure probability is small. Also, as the confusability becomes larger, this region moves to the place in which the failure probability is large. In the case of equal priors, for instance, the region without the enhancement is $0.145\le Q\le 0.631$ in Fig.~\ref{fig:3}(a), $0.245\le Q\le0.7736$ in Fig.~\ref{fig:3}(b), and $0.288\le Q\le 0.894$ in Fig.~\ref{fig:3}(c).

\subsection{Generalization to mixed state discrimination}
{For practical relevance, we further extend the aforementioned discussion to the case of discriminating two mixed qubit states. We particularly consider the equal prior probabilities to mainly focus on relation between the enhancement and the noise strength. It is assumed that the following two noisy qubit states are discriminated with the fixed failure probability~\cite{m.a.nielson}:
\begin{eqnarray}\label{noisy}
    \hat{\rho}_x=\epsilon |\psi_x\rangle\langle\psi_x|+(1-\epsilon)\frac{\hat{\mathbb{I}}_2}{2},
\end{eqnarray}
where $\epsilon\in[0,1)$ denotes the noise strength, and $\hat{\mathbb{I}}_2$ denotes an identity operator. In this case, the optimization problem of Eq.~(\ref{opt_prob}) is generalized to
\begin{eqnarray}
        \mathrm{maximize} & \  & P_{\rm succ.}^{\rm (Q)}=\frac{1}{2}\mathrm{Tr}\left\{\hat{\rho}_1\hat{M}_1\right\}+\frac{1}{2}\mathrm{Tr}\left\{\hat{\rho}_2\hat{M}_2\right\} \nonumber\\
    \mathrm{subject \ to} & \ & \hat{M}_y\ge0, \ \hat{M}_0+\hat{M}_1+\hat{M}_2=\mathbb{I}, \nonumber\\
    && Q=\mathrm{Tr}\left\{\hat{\rho}\hat{M}_0\right\}=\mathrm{constant},
\end{eqnarray}
with $\hat{\rho}=\frac{1}{2}\hat{\rho}_1+\frac{1}{2}\hat{\rho}_2$. It is shown that an optimal measurement has a balanced structure such that  $\langle\psi_1|\hat{M}_0|\psi_1\rangle=\langle\psi_2|\hat{M}_0|\psi_2\rangle$ and $\langle\psi_1^\bot|\hat{M}_0|\psi_1^\bot\rangle=\langle\psi_2^\bot|\hat{M}_0|\psi_2^\bot\rangle$~\cite{j.fiurasek}. }

\begin{figure}[t]
    \centerline{\includegraphics[width=8cm]{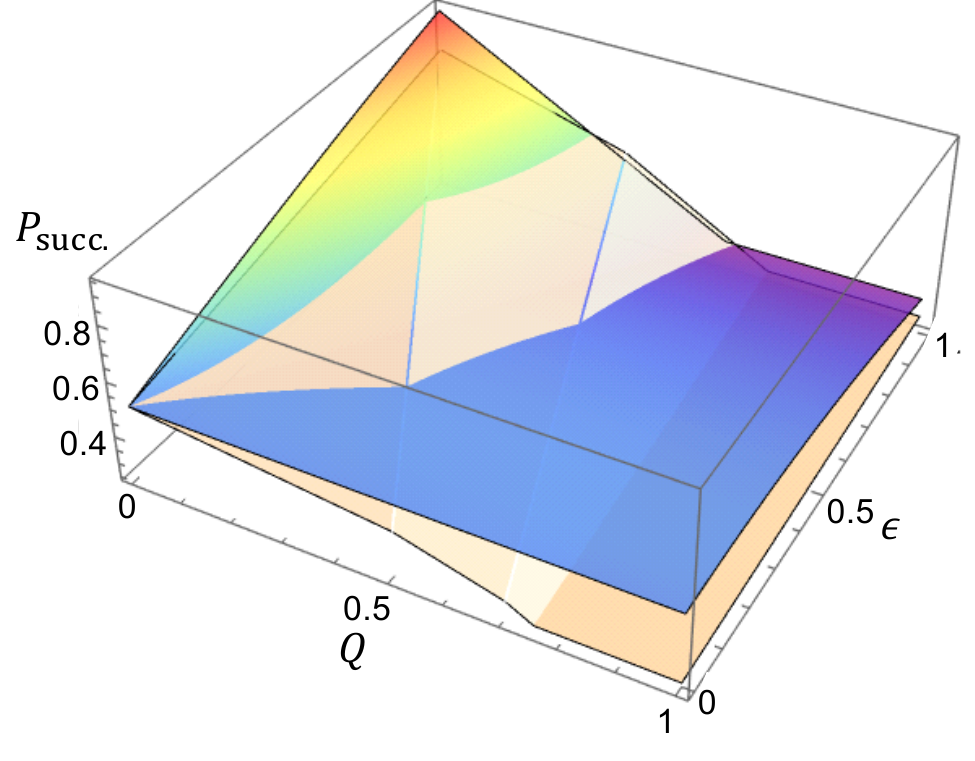}}
    \caption{{Contextuality-enhanced success probability investigated across $0\le Q,\epsilon\le 1$ for given confusability $c_{\psi_1,\psi_2}=0.4$. Here, the surface colored in rainbow is the maximum success probability under the quantum model, and the orange surface is the noncontextual bound.} }
    \centering
    \label{fig:4}
    \end{figure}

{We verify that the balanced measurement can exhibit the contextuality enhancement. In the noncontextual model, the noisy qubit states of Eq.~(\ref{noisy}) are represented by the probability density function of the hidden variable:
\begin{eqnarray}
    p_\rho(\lambda|x)=\epsilon p_{\psi}(\lambda|x)+(1-\epsilon)\frac{p_{\psi}(\lambda|x)+p_{\psi^{\bot}}(\lambda|x)}{2},
\end{eqnarray}
where $p_\psi(\lambda|x)$ and $p_{\psi^\bot}(\lambda|x)$ are associated with a pure state $|\psi_x\rangle$ in Eq.~(\ref{noisy}) and $|\psi_x^\bot\rangle$ that is orthogonal to $|\psi_x\rangle$, respectively. The above modeling is rational in the aspect that the identity operator in Eq.~(\ref{noisy}) can have the spectral decomposition with $|\psi_x\rangle$ and $|\psi_x^\bot\rangle$. In the same way that we conducted in Sec. III-A, we derive the following theorem regarding the noncontextual bound (The detailed derivation is provided in Appendix C).}

{\textit{Theorem 2.} When a balanced measurement is employed for discriminating mixed states, the success probability of the noncontextual model is upper-bounded by
\begin{eqnarray}\label{mix_an}
    \bar{P}_{\rm succ.}^{\rm (NC)}(Q)=\begin{cases}
        \frac{1+(1-c_{\psi_1,\psi_2})\epsilon-c_{\psi_1,\psi_2}Q}{2}  \ \ \mathrm{if} \  Q\le \frac{1-\epsilon}{2}, \\ \frac{3-c_{\psi_1,\psi_2}+(1-c_{\psi_1,\psi_2})\epsilon-2Q}{4} \\ \ \ \ \ \ \  \mathrm{if} \ \frac{1-\epsilon}{2}\le Q\le\frac{1-\epsilon+c_{\psi_1,\psi_2}(1+\epsilon)}{2}, \\
        1-Q \ \mathrm{otherwise,}
    \end{cases}
\end{eqnarray}
for a given $Q$ and a noise strength $\epsilon$.}

{It is noted that the maximum success probability is $1-Q_{\rm max}$ as $Q$ is larger than $Q_{\rm max}$, allowed by the maximal-confidence discrimination. This leads to the success probability~\cite{k.flatt2}
\begin{eqnarray}
    P_{\rm succ.}^{\rm (NC)}=1-Q_{\rm max}^{\rm (NC)}=\frac{1}{2}\left(1-\frac{1-\epsilon}{2}-\epsilon c_{\psi_1,\psi_2}\right).
\end{eqnarray}}

{Figure~\ref{fig:4} illustrates the contextuality enhancement in discriminating two noisy qubit states. In Fig.~\ref{fig:4}, we observe that the contextuality enhancement is possible when the fixed failure probability $Q$ is not included in the intermediate region. Moreover, this intermediate region tends to reduce with increasing $\epsilon$, suggesting the pronounced enhancement.}

\section{Conclusion}
In this work, we studied the contextuality enhancement in the quantum state discrimination strategy where the failure probability is fixed to a constant. We first established the noncontextual bound--an upper limit of the success probability under the noncontextual hidden-variable model--which applies to arbitrary fixed failure probabilities, prior probabilities, and confusability between states. Using this bound, we theoretically verified the occurrence of contextuality-enhanced success probability in terms of the failure probability. We found that contextuality enhancement does not always occur; rather, it is absent within a specific intermediate region of the failure probability. This result highlights the importance of identifying such a region when generalizing state discrimination strategies beyond minimum-error and unambiguous discrimination. Furthermore, we demonstrated that the boundary of this region is governed by the confusability between the quantum states, which quantifies their indistinguishability. As confusability increases, the enhancement tends to emerge only at higher failure probabilities; conversely, for less confusable states, enhancement is present even at lower failure probabilities. {Finally, we extended our framework to the discrimination of two mixed quantum states. In such cases, the fixed-failure probability strategy connects the minimum-error and maximum-confidence discrimination. We propose an interesting behavior that the contextual aspect of the state discrimination tends to increase along with the noise strength.}

We believe that our investigation contributes to the deeper understanding of quantum contextuality and paves the way for its practical application in quantum information science and technology. Notably, both minimum-error and unambiguous state discrimination have been employed in various quantum information tasks such as quantum-enhanced communication~\cite{i.a.burenkov,g.cariolaro}, quantum cloning~\cite{v.buzek,l.-m.duan,m.lostaglio}, and randomness certification~\cite{j.b.brask,c.r.i.carceller}. {Specifically, the state discrimination scheme considered in our work applies to quantum communication based on binary coherent states. Both minimum-error and unambiguous discrimination schemes assume perfect photodetectors~\cite{s.j.dolinar,k.banaszek}, whereas the present scheme accommodates realistic detectors with inefficiency and dark counts that result in failure outcomes~\cite{g.cariolaro}.} Our methodology can be particularly relevant to experimental scenarios where both error and failure events naturally arise. {This also suggests that our results provide a practical guide for testing the nonclassicality of a single quantum system even with imperfect measurements, while circumventing the {non-enhancement} cases.} Finally, it would be interesting to extend this framework to the discrimination of three or more mixed quantum states{, potentially leading to a unified verification of contextuality across a broader class of state discrimination strategies, including those in the presence of noise.}

\begin{widetext}
\section*{Appendix A. Detailed derivationn of noncontextual bound}
In this section, we derive the theoretical bound of the success probability in detail. Let us first begin this by rewriting the success probability in the noncontextual model:
\begin{align}\label{p_nc_b}
    P_{\rm succ.}^{\rm (NC)}&=q_1\int_{\Omega}d\lambda p_\psi (\lambda|1)p_{\mathcal{M}}(1|\lambda)+q_2\int_{\Omega}d\lambda p_\psi (\lambda|2)p_{\mathcal{M}}(2|\lambda)\nonumber\\
    &=q_1\int_{\mathrm{supp}[p_{\mathcal{M}}(1|\lambda)]}d\lambda p_\psi (\lambda|1)p_{\mathcal{M}}(1|\lambda)+q_2\int_{\mathrm{supp}[p_{\mathcal{M}}(2|\lambda)]}d\lambda p_\psi (\lambda|2)p_{\mathcal{M}}(2|\lambda)\nonumber\\
    &{=}  q_1\int_{\mathrm{supp}[p_{\mathcal{M}}(1|\lambda)]\cup\mathrm{supp}[p_{\mathcal{M}}(2|\lambda)]}d\lambda p_\psi (\lambda|1)p_{\mathcal{M}}(1|\lambda)+q_2\int_{\mathrm{supp}[p_{\mathcal{M}}(1|\lambda)]\cup\mathrm{supp}[p_{\mathcal{M}}(2|\lambda)]}d\lambda p_\psi (\lambda|2)p_{\mathcal{M}}(2|\lambda).
\end{align}
The first equality holds since $p_{\mathcal{M}}(x|\lambda)$ is strictly positive for $\lambda\in\mathrm{supp}[p_{\mathcal{M}}(x|\lambda)]$ and zero otherwise. The second {equality} is proven using {the following relations:
\begin{align}
    &\int_{\mathrm{supp}[p_{\mathcal{M}}(1|\lambda)]\cup\mathrm{supp}[p_{\mathcal{M}}(2|\lambda)]}d\lambda p_\psi (\lambda|x)p_{\mathcal{M}}(x|\lambda)\nonumber\\
    &=\int_{\mathrm{supp}[p_{\mathcal{M}}(x|\lambda)]}d\lambda p_\psi (\lambda|x)p_{\mathcal{M}}(x|\lambda)+\int_{\mathrm{supp}[p_{\mathcal{M}}(x|\lambda)]^{\rm c}\cap\mathrm{supp}[p_{\mathcal{M}}(x\oplus1|\lambda)]}d\lambda p_\psi (\lambda|x)p_{\mathcal{M}}(x|\lambda)=\int_{\mathrm{supp}[p_{\mathcal{M}}(x|\lambda)]}d\lambda p_\psi (\lambda|x)p_{\mathcal{M}}(x|\lambda),
\end{align}
with $x\oplus1=(x\mathrm{ \ mod \ }2)+1$. In the above relations: the first equality is a straightforward consequence of set-theoretic considerations; and the second equality is verified from that $p_{\mathcal{M}}(x|\lambda)$ is zero for any $\lambda\in\mathrm{supp}[p_{\mathcal{M}}(x|\lambda)]^{\rm c}\cap\mathrm{supp}[p_{\mathcal{M}}(x\oplus1|\lambda)]$.} It is obvious that a function $\Pi(\lambda)=p_{\mathcal{M}}(1|\lambda)+p_{\mathcal{M}}(2|\lambda)$ has its inverse $\Pi^{-1}(\lambda)=\frac{1}{\Pi(\lambda)}$ if $\lambda\in\mathrm{supp}[p_{\mathcal{M}}(1|\lambda)]\cup\mathrm{supp}[p_{\mathcal{M}}(2|\lambda)]$, because either $p_{\mathcal{M}}(1|\lambda)$ and $p_{\mathcal{M}}(2|\lambda)$ is not zero in this region. This leads to the inequality
\begin{align}\label{p_nc_til}
    P_{\rm succ.}^{\rm (NC)}\le(1-Q)\left\{\frac{q_1(1-Q_1)}{1-Q}\int_{\mathrm{supp}[\Pi(\lambda)]}d\lambda p_{\widetilde{\psi}}(\lambda|1)p_{\widetilde{\mathcal{M}}}(1|\lambda)+\frac{q_2(1-Q_2)}{1-Q}\int_{\mathrm{supp}[\Pi(\lambda)]}d\lambda p_{\widetilde{\psi}}(\lambda|2)p_{\widetilde{\mathcal{M}}}(2|\lambda)\right\},
\end{align}
where $Q_x$ with $x=1,2$ and $Q$ are defined as
\begin{align}\label{Q_til}
    Q_x=\int_{\mathrm{supp}[\Pi(\lambda)]}d\lambda p_{\psi}(x|\lambda)\{1-\Pi(\lambda)\}, \ Q=q_1 Q_1+q_2 Q_2, 
\end{align}
respectively. It is noted that $\Pi(\lambda)>0$ if and only if $p_{\mathcal{M}}(1|\lambda)>0$ or $p_{\mathcal{M}}(2|\lambda)>0$, leading to $\mathrm{supp}[\Pi(\lambda)]=\mathrm{supp}[p_{\mathcal{M}}(1|\lambda)]\cup\mathrm{supp}[p_{\mathcal{M}}(2|\lambda)]$. In Eq.~(\ref{Q_til}), the probability distribution $p_{\widetilde{\psi}}(\lambda|x)$ is described as
\begin{equation}\label{psi_til}
    p_{\widetilde{\psi}}(\lambda|x)=\frac{p_{\psi}(\lambda|x)\Pi(\lambda)}{1-Q_x},
\end{equation}
and the response function $p_{\widetilde{\mathcal{M}}}(x|\lambda)$ is
\begin{equation}
    p_{\widetilde{\mathcal{M}}}(x|\lambda)=\frac{1}{\Pi(\lambda)}p_{\mathcal{M}}(x|\lambda).
\end{equation}
According to what was shown by Refs.~\cite{s.mukerjee,j.shin}, the success probability in Eq.~(\ref{p_nc_til}) is upper-bounded by
\begin{equation}\label{p_nc_b1}
    P_{\rm succ.}^{\rm (NC)}\le(1-Q)\left[1-\min\left\{\frac{q_1(1-Q_1)}{1-Q},\frac{q_2(1-Q_2)}{1-Q}\right\}c_{\widetilde{\psi}_1,\widetilde{\psi}_2}\right],
\end{equation}
where $c_{\widetilde{\psi}_1,\widetilde{\psi}_2}$ is the confusability between $p_{\widetilde{\psi}}(1|\lambda)$ and $p_{\widetilde{\psi}}(2|\lambda)$, defined as
\begin{equation}
    c_{\widetilde{\psi}_1,\widetilde{\psi}_2}=\int_{\mathrm{supp}[p_{\widetilde{\psi}}(1|\lambda)]}d\lambda p_{\widetilde{\psi}}(2|\lambda)=\int_{\mathrm{supp}[p_{\widetilde{\psi}}(2|\lambda)]}d\lambda p_{\widetilde{\psi}}(1|\lambda).
\end{equation}
Together with Eq.~(\ref{psi_til}), we can show that the confusability $c_{\widetilde{\psi}_1,\widetilde{\psi}_2}$ satisfies
\begin{align}
    c_{\widetilde{\psi}_1,\widetilde{\psi}_2}&=\int_{\mathrm{supp}[p_\psi(2|\lambda)]}d\lambda\frac{p_\psi(1|\lambda)\Pi(\lambda)}{1-Q_1}\nonumber\\
    &=\frac{1}{1-Q_1}\left[\int_{\mathrm{supp}[p_\psi(2|\lambda)]}d\lambda p_\psi(1|\lambda)-\int_{\mathrm{supp}[p_\psi(2|\lambda)]}d\lambda p_\psi(1|\lambda)p_{\mathcal{M}}(\lambda|0)\right]\nonumber\\
    &\ge\frac{c_{\psi_1,\psi_2}-Q_1}{1-Q_1}.
\end{align}
Here the first equality is satisfied since $\int_{\mathrm{supp}[p_\psi(2|\lambda)]\cap\mathrm{ker}[\Pi(\lambda)]}d\lambda{p_\psi(1|\lambda)\Pi(\lambda)}$, in which $\mathrm{ker}[\Pi(\lambda)]:=\{\lambda\in\Omega:\Pi(\lambda)=0\}$, is zero and thus $\int_{\mathrm{supp}[p_{\widetilde{\psi}}(2|\lambda)]}d\lambda{p_\psi(1|\lambda)\Pi(\lambda)}=\int_{\mathrm{supp}[p_\psi(2|\lambda)]}d\lambda{p_\psi(1|\lambda)\Pi(\lambda)}$, the second equality is straightforward due to $\Pi(\lambda)=1-p_{\mathcal{M}}(0|\lambda)$ $\forall\lambda\in\Omega$, and the third inequality is derived using the fact that $\mathrm{supp}[p_\psi(2|\lambda)]\subseteq\Omega$. {It is further noted that the third inequality becomes an equality when the support space of $p_{\mathcal{M}}(\lambda|0)$ covers $\mathrm{supp}[p_{\psi}(1|\lambda)]\cap\mathrm{supp}[p_{\psi}(2|\lambda)]$.} In the same manner, the confusability $c_{\widetilde{\psi}_1,\widetilde{\psi}_2}$ is lower-bounded by $\frac{c_{\psi_1,\psi_2}-Q_2}{1-Q_2}$. Consequently, we obtain the following inequality
\begin{equation}\label{c_bound}
    c_{\widetilde{\psi}_1,\widetilde{\psi}_2}\ge\max\left\{\frac{c_{\psi_1,\psi_2}-Q_1}{1-Q_1},\frac{c_{\psi_1,\psi_2}-Q_2}{1-Q_2},0\right\}.
\end{equation}
Substituting the above inequality to Eq.~(\ref{p_nc_b1}) yields
\begin{equation}\label{p_nc_b2}
    P_{\rm succ.}^{\rm (NC)}\le 1-Q-\min\left\{q_1(1-Q_1),q_2(1-Q_2)\right\}\max\left\{\frac{c_{\psi_1,\psi_2}-Q_1}{1-Q_1},\frac{c_{\psi_1,\psi_2}-Q_2}{1-Q_2},0\right\},
\end{equation}
which is admitted for all possible $Q_1$ and $Q_2$ satisfying $0\le Q_1,Q_2\le 1$ and $Q=q_1Q_1+q_2Q_2$. Finally, we obtain the upper bound $\bar{P}_{\rm succ.}^{\rm (NC)}$  as
\begin{equation}\label{p_nc_bound}
    {P}_{\rm succ.}^{\rm (NC)}\le\max_{\substack{0\le Q_1,Q_2\le 1 \\ Q=q_1Q_1+q_2Q_2}}\left[1-Q-\min\left\{q_1(1-Q_1),q_2(1-Q_2)\right\}\max\left\{\frac{c_{\psi_1,\psi_2}-Q_1}{1-Q_1},\frac{c_{\psi_1,\psi_2}-Q_2}{1-Q_2},0\right\}\right].
\end{equation}

\section*{Appendix B. Explicit form of noncontextual bound}
We derive the explicit form of the noncontextual bound by solving the optimization problem in Eq.~(\ref{p_nc_bound}). We first consider the three cases as detailed below:
\begin{itemize}
    \item Let $Q_1,Q_2\ge c_{\psi_1,\psi_2}$. Then, the bound  of Eq.~(\ref{p_nc_bound}) is simplified to ${P}_{\rm succ.}^{\rm (NC)}\le1-Q$.
    \item Let $Q_1\ge c_{\psi_1,\psi_2}$ and $Q_2\le c_{\psi_1,\psi_2}$. Then, the upper bound is rewritten as
    \begin{equation}
        {P}_{\rm succ.}^{\rm (NC)}\le\max_{\substack{0\le Q_1\le c_{\psi_1,\psi_2}\le Q_2\le 1 \\ Q=q_1Q_1+q_2Q_2}}\left[1-Q-\min\left\{q_1(1-Q_1),q_2(1-Q_2)\right\}\frac{c_{\psi_1,\psi_2}-Q_1}{1-Q_1}\right].
    \end{equation}
    In the case of $q_1(1-Q_1)\le q_2(1-Q_2)$, the above bound is simplified as
    \begin{equation}
        {P}_{\rm succ.}^{\rm (NC)}\le\max_{\substack{0\le Q_1\le c_{\psi_1,\psi_2}\le Q_2\le 1 \\ Q=q_1Q_1+q_2Q_2}}[1-Q-q_1(c_{\psi_1,\psi_2}-Q_1)]\le\max_{0\le Q_1\le c_{\psi_1,\psi_2}}[1-Q-q_1(c_{\psi_1,\psi_2}-Q_1)]=1-Q.
    \end{equation}
    Otherwise, the upper bound is written as
    \begin{align}
        {P}_{\rm succ.}^{\rm (NC)}&\le\max_{\substack{0\le Q_1\le c_{\psi_1,\psi_2}\le Q_2\le 1 \\ Q=q_1Q_1+q_2Q_2}}\left[1-Q-q_2(1-Q_2)\frac{c_{\psi_1,\psi_2}-Q_1}{1-Q_1}\right]\nonumber\\
        &\le \max_{0\le Q_1\le c_{\psi_1,\psi_2}\le Q_2\le 1}\left[1-Q-q_2(1-Q_2)\frac{c_{\psi_1,\psi_2}-Q_1}{1-Q_1}\right].
    \end{align}
    It is noted that $\frac{c_{\psi_1,\psi_2}-Q_1}{1-Q_1}$ monotonically decreases with $Q_1$ in the region of $0\le Q_1\le c_{\psi_1,\psi_2}$. Thus, the upper bound is evaluated as ${P}_{\rm succ.}^{\rm (NC)}\le1-Q$ achieved by $Q_1=c_{\psi_1,\psi_2}$.
    \item Let $Q_1\le c_{\psi_1,\psi_2}$ and $Q_2\ge c_{\psi_1,\psi_2}$. Then, the upper bound is calculated as ${P}_{\rm succ.}^{\rm (NC)}\le1-Q$, in the same way that we calculated in the above case. 
\end{itemize}
All these three cases yield the noncontextual bound $\bar{P}_{\rm succ.}^{\rm (NC)}=1-Q$. 

Now we focus on the case of $Q_1,Q_2\le c_{\psi_1,\psi_2}$. In this case, the upper bound $\bar{P}_{\rm succ.}^{\rm (NC)}$ is described from Eq.~(\ref{p_nc_bound}) as
    \begin{equation}
        \bar{P}_{\rm succ.}^{\rm (NC)}=\max_{\substack{0\le Q_1,Q_2\le c_{\psi_1,\psi_2} \\ Q=q_1Q_1+q_2Q_2}}\left[1-Q-\min\left\{q_1(1-Q_1),q_2(1-Q_2)\right\}\max\left\{\frac{c_{\psi_1,\psi_2}-Q_1}{1-Q_1},\frac{c_{\psi_1,\psi_2}-Q_2}{1-Q_2}\right\}\right].
    \end{equation}
    It is straightforward that $\bar{P}_{\rm succ.}^{\rm (NC)}=1-Q$ if $q_1(1-Q_1)\le q_2(1-Q_2)$ and $\frac{c_{\psi_1,\psi_2}-Q_1}{1-Q_1}\ge\frac{c_{\psi_1,\psi_2}-Q_2}{1-Q_2}$, or if $q_1(1-Q_1)\ge q_2(1-Q_2)$ and $\frac{c_{\psi_1,\psi_2}-Q_1}{1-Q_1}\le\frac{c_{\psi_1,\psi_2}-Q_2}{1-Q_2}$. Otherwise, we can have a nontrivial upper bound of the success probability. In the case of $q_1(1-Q_1)\le q_2(1-Q_2)$ and $\frac{c_{\psi_1,\psi_2}-Q_1}{1-Q_1}\le\frac{c_{\psi_1,\psi_2}-Q_2}{1-Q_2}$, for instance, $\bar{P}_{\rm succ.}^{\rm (NC)}$ is rewritten as
    \begin{equation}\label{p_nc_nont}
        \bar{P}_{\rm succ.}^{\rm (NC)}=\max_{Q_{\rm 1,min}\le Q_1\le Q_{\rm 1,max}}\left[1-Q-q_1(1-Q_1)\frac{q_2c_{\psi_1,\psi_2}-Q+q_1Q_1}{q_2-Q+q_1Q_1}\right]:=\max_{Q_{\rm 1,min}\le Q_1\le Q_{\rm 1,max}}f_1(Q_1),
    \end{equation}
    where $Q_{\rm 1,min}$ and $Q_{\rm 1,max}$ are
    \begin{align}\label{qb_}
        Q_{\rm 1,min}=\max\left\{0,\frac{Q-q_2c_{\psi_1,\psi_2}}{q_1}\right\}, \ \ Q_{\rm 1,max}=\min\left\{\frac{Q}{q_1},c_{\psi_1,\psi_2}\right\},
    \end{align}
    respectively. Both Eqs.~(\ref{p_nc_nont}) and (\ref{qb_}) are derived from that both $Q_1$ and $Q_2=\frac{Q-q_1Q_1}{q_2}$ should satisfy $0\le Q_x\le c_{\psi_1,\psi_2}$. To maximize $f_1(Q_1)$ in Eq.~(\ref{p_nc_nont}), let us consider two extremum points 
    \begin{equation}
        Q_{\rm 1,ext}^{(\pm)}=\frac{Q-q_2\pm\sqrt{(1-c_{\psi_1,\psi_2})(1-Q)q_2}}{q_1}.
    \end{equation}
    Since $\frac{\partial^2f_1}{\partial Q_1^{{2}}}\Big|_{Q_1=Q_{\rm 1,ext}^{(+)}}$ is positive-definite, meaning that $Q_{\rm 1,ext}^{(+)}$ is a local minimum of $f(Q_1)$, it is not a global maximum. Let us then focus on $Q_{\rm 1,ext}^{(-)}$, yielding $\frac{\partial^2f_1}{\partial Q_1^{{2}}}\Big|_{Q_1=Q_{\rm 1,ext}^{(-)}}<0$. Noting that 
    \begin{equation}
        Q_1\ge 0 \ \ \rightarrow \ \ \frac{q_2c_{\psi_1,\psi_2}-Q+q_1Q_1}{q_2-Q+q_1Q_1}\le 1,
    \end{equation}
    which is equivalent to 
    \begin{equation}
        \frac{q_2c_{\psi_1,\psi_2}-Q+q_1Q_1}{q_2-Q+q_1Q_1}> 1 \ \ \rightarrow \ \ Q_1<0,
    \end{equation}
    $Q_{\rm 1,ext}^{(-)}$ is less than zero since
    \begin{equation}
        \frac{q_2c_{\psi_1,\psi_2}-Q+q_1Q_{\rm 1,ext}^{(-)}}{q_2-Q+q_1Q_{\rm 1,ext}^{(-)}}=\frac{q_2(1-c_{\psi_1,\psi_2})+\sqrt{(1-c_{\psi_1,\psi_2})(1-Q)q_2}}{\sqrt{(1-c_{\psi_1,\psi_2})(1-Q)q_2}}>1.
    \end{equation}
    All these arguments lead to that there is no local minimum and maximum in the region of $Q_{\rm 1,min}\le Q_1\le Q_{\rm 1,max}$, implying that either $Q_{\rm 1,min}$ or $Q_{\rm 1,max}$ is the global maximum. Likewise, we derive $\bar{P}_{\rm succ.}^{\rm (NC)}$ as
    \begin{equation}
        \bar{P}_{\rm succ.}^{\rm (NC)}=\max_{Q_{\rm 2,min}\le Q_2\le Q_{\rm 2,max}}\left[1-Q-q_2(1-Q_2)\frac{q_1c_{\psi_1,\psi_2}-Q+q_2Q_2}{q_1-Q+q_2Q_2}\right]:=\max_{Q_{\rm 2,min}\le Q_2\le Q_{\rm 2,max}}f_2(Q_2),
    \end{equation}
    with
    \begin{align}\label{qb}
        Q_{\rm 2,min}=\max\left\{0,\frac{Q-q_1c_{\psi_1,\psi_2}}{q_2}\right\}, \ \ Q_{\rm 2,max}=\min\left\{\frac{Q}{q_2},c_{\psi_1,\psi_2}\right\},
    \end{align}
    where $f_2(Q_2)$ is maximized when $Q_2$ is either $Q_{\rm 2,min}$ or $Q_{\rm 2,max}$. Finally, $\bar{P}_{\rm succ.}^{\rm (NC)}$ takes the form of
    \begin{equation}
        \bar{P}_{\rm succ.}^{\rm (NC)}=\max\left\{f_1(Q_{\rm 1,min}),f_1(Q_{\rm 1,max}),f_2(Q_{\rm 2,min}),f_2(Q_{\rm 2,max})\right\},
    \end{equation}
    equivalent to Eq.~(\ref{p_nc_bf}) of Sec. III-B.

    \section*{Appendix C. Extension to mixed state discrimination}
    {We extend the derivation introduced in Appendices A and B to discrimination between two mixed qubit states of Eq.~(\ref{noisy}), represented by the noncontextual model as
    \begin{eqnarray}
        p_\rho(\lambda|x)=\epsilon p_{\psi}(\lambda|x)+(1-\epsilon)\frac{p_{\psi}(\lambda|x)+p_{\psi^{\bot}}(\lambda|x)}{2}=\frac{1+\epsilon}{2}p_{\psi}(\lambda|x)+\frac{1-\epsilon}{2}p_{\psi^{\bot}}(\lambda|x),
    \end{eqnarray}
    with $0\le\epsilon\le1$. Here, $p_{\psi^{\bot}}(\lambda|x)$ is the state orthogonal to $p_{\psi}(\lambda|x)$ in terms of the zero confusability: $\int_{\mathrm{supp}[p_{\psi}(\lambda|x)]}d\lambda p_{\psi^\bot}(\lambda|x)=0$. The success probability of the discrimination is given by
    \begin{eqnarray}\label{p_nc_mix}
        P_{\rm succ.}^{\rm (NC)}&=&q_1\int_{\mathrm{supp}[\Pi(\lambda)]}d\lambda p_\rho (\lambda|1)p_{\mathcal{M}}(1|\lambda)+q_2\int_{\mathrm{supp}[\Pi(\lambda)]}d\lambda p_\rho (\lambda|2)p_{\mathcal{M}}(2|\lambda)\\
        &=&(1-Q)\left\{\frac{q_1(1-Q_1^{(\rho)})}{1-Q}\int_{\mathrm{supp}[\Pi(\lambda)]}d\lambda p_{\widetilde{\rho}}(\lambda|1)p_{\widetilde{\mathcal{M}}}(1|\lambda)+\frac{q_2(1-Q_2^{(\rho)})}{1-Q}\int_{\mathrm{supp}[\Pi(\lambda)]}d\lambda p_{\widetilde{\rho}}(\lambda|2)p_{\widetilde{\mathcal{M}}}(2|\lambda)\right\},\nonumber
    \end{eqnarray}
    where $Q_x^{(\rho)}$ and $p_{\widetilde{\rho}}(\lambda|x)$ are defined as
    \begin{eqnarray}
        Q_x^{(\rho)}=1-\int_{\mathrm{supp}[\Pi(\lambda)]}d\lambda p_\rho (\lambda|x)\Pi(\lambda), \ p_{\rho}(\lambda|x)=\frac{p_\rho (\lambda|x)\Pi(\lambda)}{\int_{\mathrm{supp}[\Pi(\lambda)]}d\lambda p_\rho (\lambda|x)\Pi(\lambda)}=\frac{p_\rho (\lambda|x)\Pi(\lambda)}{1-Q_x^{(\rho)}},
    \end{eqnarray}
    respectively. It is straightforward that the success probability of Eq.~(\ref{p_nc_mix}) is rewritten as
    \begin{align}\label{succ_nc_2}
        P_{\rm succ.}^{\rm (NC)}=(1-Q)\Bigg[&\frac{q_1(1-Q_1^{(\rho)})}{1-Q}\int_{\mathrm{supp}[\Pi(\lambda)]}d\lambda\left\{\frac{1+\epsilon}{2}\frac{1-Q_1^{(\psi)}}{1-Q_1^{(\rho)}}p_{\widetilde{\psi}}(\lambda|1)+\frac{1-\epsilon}{2}\frac{1-Q_1^{(\psi^\bot)}}{1-Q_1^{(\rho)}}p_{\widetilde{\psi}^{\bot}}(\lambda|1)\right\}p_{\widetilde{\mathcal{M}}}(1|\lambda)\nonumber\\
        +&\frac{q_2(1-Q_2^{(\rho)})}{1-Q}\int_{\mathrm{supp}[\Pi(\lambda)]}d\lambda\left\{\frac{1+\epsilon}{2}\frac{1-Q_2^{(\psi)}}{1-Q_2^{(\rho)}}p_{\widetilde{\psi}}(\lambda|2)+\frac{1-\epsilon}{2}\frac{1-Q_2^{(\psi^\bot)}}{1-Q_2^{(\rho)}}p_{\widetilde{\psi}^{\bot}}(\lambda|2)\right\}p_{\widetilde{\mathcal{M}}}(2|\lambda)\Bigg],
    \end{align}
    where $p_{\widetilde{\psi}}(\lambda|x)$ and $p_{\widetilde{\psi}^\bot}(\lambda|x)$ denote
    \begin{eqnarray}
        p_{\widetilde{\psi}}(\lambda|x)&=&\frac{p_\psi(\lambda|x)\Pi(\lambda)}{\int_{\mathrm{supp}[\Pi(\lambda)]}p_\psi(\lambda|x)\Pi(\lambda)}=\frac{p_\psi(\lambda|x)\Pi(\lambda)}{1-Q_x^{(\psi)}},\nonumber\\
        p_{\widetilde{\psi}^\bot}(\lambda|x)&=&\frac{p_{\psi^\bot}(\lambda|x)\Pi(\lambda)}{\int_{\mathrm{supp}[\Pi(\lambda)]}p_{\psi^\bot}(\lambda|x)\Pi(\lambda)}=\frac{p_{\psi^\bot}(\lambda|x)\Pi(\lambda)}{1-Q_x^{(\psi^\bot)}},
    \end{eqnarray}
    respectively. Both $p_{\widetilde{\psi}}(\lambda|x)$ and $p_{\widetilde{\psi}^\bot}(\lambda|x)$ are orthogonal to each other, as verified as
    \begin{eqnarray}
        \int_{\mathrm{supp}[p_{\widetilde{\psi}^\bot}(\lambda|x)]}d\lambda p_{\widetilde{\psi}}(\lambda|x)\propto\int_{\mathrm{supp}[p_{{\psi}^\bot}(\lambda|x)]\cap\mathrm{supp}[\Pi(\lambda)]}d\lambda p_{\psi}(\lambda|x)\Pi(\lambda)=0,
    \end{eqnarray}
    together with a simple set-theoretic identities
    \begin{eqnarray}
        \left\{\mathrm{supp}[p_{{\psi}^\bot}(\lambda|x)]\cap\mathrm{supp}[\Pi(\lambda)]\right\}\cap\mathrm{supp}[p_{{\psi}}(\lambda|x)]=\left\{\mathrm{supp}[p_{{\psi}^\bot}(\lambda|x)]\cap\mathrm{supp}[p_{{\psi}}(\lambda|x)]\right\}\cap\mathrm{supp}[\Pi(\lambda)]=\phi.
    \end{eqnarray}
    We assume the equal prior probabilities ($q_1=q_2$) to focus on the effect of the noise. It is rational to simplify $Q_x^{(\psi)}=Q^{(\psi)}$, $Q_x^{(\psi^\bot)}=Q^{(\psi^\bot)}$, and $Q_x^{(\rho)}=Q^{(\rho)}$ for any $x=1,2$ when considering the balanced measurement. Then, the success probability of Eq.~(\ref{succ_nc_2}) is simplified to
    \begin{align}
        P_{\rm succ.}^{\rm (NC)}=(1-Q)\Bigg[&\frac{q_1(1-Q^{(\rho)})}{1-Q}\int_{\mathrm{supp}[\Pi(\lambda)]}d\lambda\left\{\frac{1+f}{2}p_{\widetilde{\psi}}(\lambda|1)+\frac{1-f}{2}p_{\widetilde{\psi}^{\bot}}(\lambda|1)\right\}p_{\widetilde{\mathcal{M}}}(1|\lambda)\nonumber\\
        +& \frac{q_2(1-Q^{(\rho)})}{1-Q}\int_{\mathrm{supp}[\Pi(\lambda)]}d\lambda\left\{\frac{1+f}{2}p_{\widetilde{\psi}}(\lambda|2)+\frac{1-f}{2}p_{\widetilde{\psi}^{\bot}}(\lambda|2)\right\}p_{\widetilde{\mathcal{M}}}(2|\lambda)\Bigg],
    \end{align}
    where $f$ is the parameter defined as
    \begin{eqnarray}\label{ff}
        f=\frac{1+\epsilon}{2}\frac{1-Q^{(\psi)}}{1-Q^{(\rho)}}-\frac{1-\epsilon}{2}\frac{1-Q^{(\psi^\bot)}}{1-Q^{(\rho)}}.
    \end{eqnarray}
    Now, we can use Eq. (22) of the previous work~\cite{j.shin}, together with Eqs.~(\ref{c_bound}) and (\ref{ff}), to derive the inequality
    \begin{eqnarray}
        P_{\rm succ.}^{\rm (NC)}&\le&(1-Q)\left\{\frac{1+f}{2}-\frac{f}{2}c_{\widetilde{\psi}_1,\widetilde{\psi}_2}\right\}\\
        &=&\frac{1+\epsilon}{2}(1-Q^{(\psi)})-\left\{\frac{1+\epsilon}{4}(1-Q^{(\psi)})-\frac{1-\epsilon}{4}(1-Q^{(\psi^\bot)})\right\}\max\left\{\frac{c_{\psi_1,\psi_2}-Q^{(\psi)}}{1-Q^{(\psi)}},0\right\}. \nonumber
    \end{eqnarray}
    The above inequality is satisfied by any $Q^{(\psi)}$ and $Q^{(\psi^\bot)}$, consequently leading to the upper bound of the success probability
    \begin{align}
        \bar{P}_{\rm succ.}^{\rm (NC)}=\max_{\substack{0\le Q^{(\psi)},Q^{(\psi^\bot)}\le1 \\ Q=\frac{1+\epsilon}{2}Q^{(\psi)}+\frac{1-\epsilon}{2}Q^{(\psi^\bot)}}}\left[\frac{1+\epsilon}{2}(1-Q^{(\psi)})-\left\{\frac{1+\epsilon}{4}(1-Q^{(\psi)})-\frac{1-\epsilon}{4}(1-Q^{(\psi^\bot)})\right\}\max\left\{\frac{c_{\psi_1,\psi_2}-Q^{(\psi)}}{1-Q^{(\psi)}},0\right\}\right],
    \end{align}
    which is algebraically calculated as written in Eq.~(\ref{mix_an}).}
\end{widetext}

\end{document}